\documentclass[aps,prd,floats,balancelastpage,showpacs,showkeys,preprintnumbers,floatfix,nofootinbib,superscriptaddress, longbibliography]{revtex4-1}
\usepackage{mathrsfs}
\usepackage{amsfonts}
\usepackage{amsmath}
\usepackage{amssymb}
\usepackage{array}
\usepackage{verbatim}
\usepackage{bm}
\usepackage{epsfig}
\usepackage{graphicx,color}
\usepackage{relsize}
\usepackage{float}
\usepackage{multirow}
\RequirePackage{xspace}
\usepackage{graphicx}
\usepackage{amsfonts}
\usepackage{hyperref}
\usepackage{xcolor}

\begin{document}

\newcommand{\EM}[1]{\textcolor{teal}{[EM: #1]}}
\newcommand{\IR}[1]{\textcolor{red}{[IR: #1]}}
\newcommand{\AV}[1]{\textcolor{orange}{[AV: #1]}}
\newcommand{\VG}[1]{\textcolor{blue}{[VG: #1]}}

\newcommand{\Rev}[1]{\textbf{\textcolor{violet}{[Review: #1]}}}

\title{Gamma-ray production in the cosmic-ray -- dark matter scattering as a probe of the axion-like particle -- proton interaction}
\author{Victor P. Goncalves}
\email{barros@ufpel.edu.br}
\affiliation{Institute of Physics and Mathematics, Federal University of Pelotas, \\
  Postal Code 354,  96010-900, Pelotas, RS, Brazil}

\author{Emmanuel Moulin}
\email{emmanuel.moulin@cea.fr}
\affiliation{Irfu, CEA Saclay, Université Paris-Saclay, F-91191 Gif-sur-Yvette, France}

\author{Igor Reis}
\email{igorreis@ifsc.usp.br}
\email{igor.reis@cea.fr}
\affiliation{Universidade de São Paulo, Instituto de Física de São Carlos, Av. Trabalhador São Carlense 400, São Carlos, Brazil}
\affiliation{Irfu, CEA Saclay, Université Paris-Saclay, F-91191 Gif-sur-Yvette, France}

\author{Aion Viana}
\email{aion.viana@ifsc.usp.br}
\affiliation{Universidade de São Paulo, Instituto de Física de São Carlos, Av. Trabalhador São Carlense 400, São Carlos, Brazil}

\begin{abstract}
The production of very-high-energy (VHE, $E_{\gamma} \gtrsim 100$ GeV) gamma rays resulting from the scattering of high-energy cosmic-ray protons off axion-like particles (ALPs) populating the dark matter halo of the Milky Way is investigated. By employing the latest instrument response functions for current and future facilities, we demonstrate that ground-based VHE gamma-ray observatories, such as H.E.S.S., CTAO, and SWGO, provide a promising and complementary avenue to probe the yet uncharted ALP-proton coupling $g_{ap}$. Our results show that these experiments can reach sensitivity to couplings above $10^{-2}$ in the $1 - 10^{8}$ eV ALP mass range, a region that remains largely unexplored by supernova and neutron star cooling observations. Interestingly, we demonstrate that this search channel is capable of probing QCD axion dark matter models, assuming two benchmark models for it: the Kim-Shifman-Vainshtein-Zakharov (KSVZ)
Dine-Fischler-Srednicki-Zhitnitsky (DFSZ) models, specifically within the MeV mass range. These findings highlight the potential of VHE gamma-ray astronomy to provide unique constraints on the interaction between ALPs and the baryonic sector.
\end{abstract}

\maketitle
\date{\today}

\section{Introduction}

The investigation of gamma-ray production resulting from the scattering of high-energy cosmic-ray (CR) protons off Dark Matter (DM) has become a well-established approach to probing the couplings and masses of particles proposed in various Beyond the Standard Model (BSM) scenarios (see, \textit{e.g.}, Refs.~\cite{Mirizzi:2007hr,Simet:2007sa,Ayala:2014pea,MillerBertolami:2014rka,Guo:2020oum,HESS:2016mib,HESS:2022ygk,Reis:2024wfy,Goncalves:2025nij}). 
This methodology provides a unique window into the DM sector, particularly for candidates that interact weakly with standard model baryons or leptons. 
Specifically, Ref.~\cite{Goncalves:2025nij} focused on scenarios where the dark matter is composed of Axion-like particles (ALPs), which are pseudo-Nambu-Goldstone bosons that naturally arise in BSM theories~\cite{PaolaArias_2012,PeterSvrcek_2006,PhysRevD.81.123530} from the breaking of a $U(1)$ symmetry (for a review, see Ref.~\cite{Irastorza:2018dyq}). That work derived new constraints on the magnitude of the  ALP couplings to photons and electrons considering the current and future astrophysical measurements of very-high-energy (VHE, E$\gtrsim$100 GeV) gamma rays. 

\begin{figure}[b]
    \centering
    \includegraphics[width=0.6\linewidth]{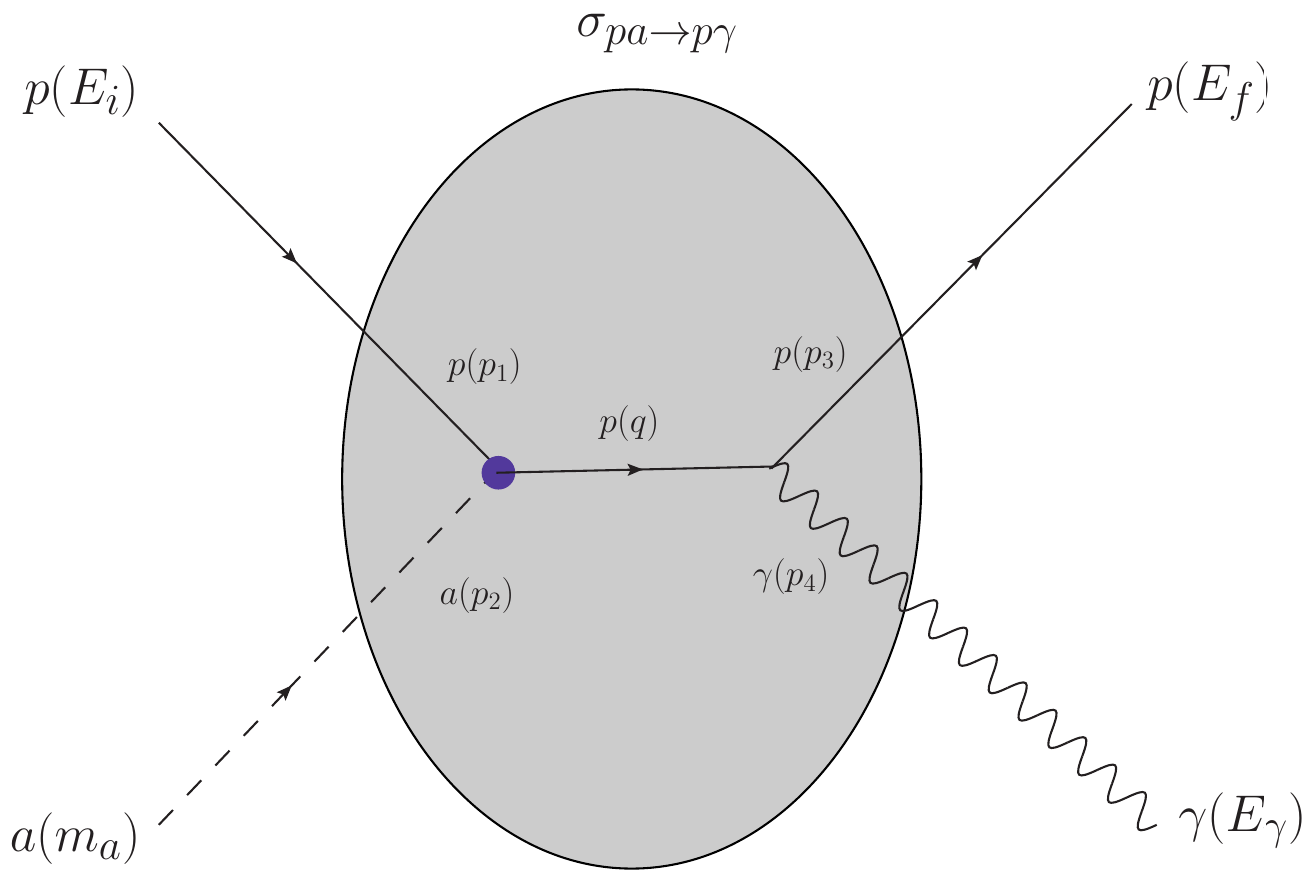}
    \caption{Gamma - ray production via ALP - proton interaction. The blue blob represents the ALP - proton coupling.}
    \label{fig:diag}
\end{figure}

In this paper, we extend the analysis performed in Ref.~\cite{Goncalves:2025nij}, by exploring the  potential to constrain the coupling between ALPs and protons ($g_{ap}$). We consider the gamma-ray production resulting from the scattering of energetic CR protons off ALPs populating the Milky Way (MW) DM halo. This process is illustrated in the Feynman diagram in Fig.~\ref{fig:diag}. An incoming CR proton of energy $E_{\rm i}$ collides with an ALP $a$ of mass $m_a$, loses a fraction of its energy, and yields a gamma ray of energy $E_{\gamma}$ in the final state. The cross-section for this process, $\sigma_{pa \rightarrow p\gamma}$, depends on the ALP mass and the magnitude of the ALP - proton coupling, represented by the blue vertex in Fig.~\ref{fig:diag}.
The associated gamma-ray flux can be calculated given the functional form of the cosmic ray flux and the MW DM distribution. Our goal is to is to evaluate the sensitivity of current and forthcoming VHE gamma-ray observatories, such as the Cherenkov Telescope Array Observatory (CTAO) \cite{CTAO} and the Southern Wide-field Gamma-ray Observatory (SWGO) \cite{SWGO}, particularly when targeting the Galactic Center (GC). As we will demonstrate, these observatories will be able to probe the $(g_{ap},m_a)$ parameter space in regions currently unexplored by supernova cooling limits, neutron star observations, or neutrino water Cherenkov detectors (for recent studies, see Refs.~\cite{Lella:2023bfb,Lella:2024dmx,Benabou:2024jlj,Chakraborty:2024tyx,Alonso-Gonzalez:2024ems,Alonso-Gonzalez:2024spi}).

The paper is organized as follows. In Section \ref{sec2} we describe gamma-ray production mechanism from CR proton - ALP scattering. In Section \ref{sec3}, we establish the methodology, both for the experimental observations of this signal, and for the statistical analysis used to derive sensitivity prospects. Finally, in Section \ref{sec4}, we present our results, comparing the reach of CTAO, SWGO and a H.E.S.S.-like observatory against existing constraints in the parameter space of the CR proton - ALP interaction. 

\section{Gamma - ray production via proton CR - ALP interactions}\label{sec2}

Initially, we briefly review of the formalism developed to 
compute the expected gamma-ray flux associated with the interaction between VHE cosmic-ray proton and the ALPs populating the Milky Way DM halo. In our analysis, we focus on photophobic ALPs, \textit{i.e.}, processes where the ALP does not couple directly to photons, making the process represented in Fig.~\ref{fig:diag} the dominant contribution. Moreover, we assume that the proton remains intact in the final state after the interaction with the ALP, \textit{i.e.}, the ALP couples to the proton as a whole and the process is described in terms of the ALP - proton coupling $g_{ap}$. For simplicity, we treat the proton as a point-like fermion. Consequently, the ALP phenomenology is determined by the effective interaction Lagrangian~\cite{Lella:2024dmx} written as
\begin{eqnarray}
    {\cal{L}}_{int} =  \frac{1}{2}\frac{g_{ap}}{m_p}(\partial_{\mu}a)\bar{p}\gamma^{\mu}\gamma_5p \,\,, 
\end{eqnarray}
where $m_p$ is the proton mass, and $a$ and $p$ represent the ALP and proton, respectively. 

In recent literature, values of the ALP - proton coupling between $10^{-9}$ and $10^{-6}$ have been excluded for ALP masses up to $\approx$ 100 MeV based on observations of supernova explosions~\cite{Lella:2023bfb,Lella:2024dmx,Benabou:2024jlj,Chakraborty:2024tyx}. More recently, the authors of Ref.\cite{Alonso-Gonzalez:2024spi} demonstrated that the parameter space region characterized by $10^{-4} \le m_a \le 1.0$ MeV and $3 \times 10^{-6} \le g_{ap} \le 4 \times 10^{-5}$ can be explored using neutrino water Cherenkov detectors, such as Super-Kamiokande \cite{Super-Kamiokande:2021jaq}. In the following, we investigate whether unexplored regions of this parameter space can be probed by the ongoing and near-future very-high-energy gamma-ray observatories.

Following Refs.~\cite{Dent:2020qev,Goncalves:2025nij}, the expected gamma-ray flux from any region of the sky within a solid angle $\Delta\Omega$ is given by: 
\begin{eqnarray}
 & \,&    \frac{d\Phi_\gamma (E_{\gamma})}{dE_{\gamma}} = \frac{1}{m_{a}} \times D(\Delta\Omega)  \times  \int_{E_{p}^{\rm min}(E_{\gamma})} dE_p \frac{d\Phi_p}{dE_p}\big(E_p\big) \cdot  \frac{d\sigma_{p+a \rightarrow p + \gamma}}{dE_{\gamma}}\big(E_p,E_\gamma\big)    \, ,
    \label{eq:flux}
\end{eqnarray}
where the $D$-factor corresponds to the integral of DM density in the MW halo along  the line-of-sight  (l.o.s.) $s$ and over $\Delta\Omega$. 
The DM density distribution is assumed to follow
the Navarro–Frenk–White (NFW) profile parametrization~\footnote{A detailed analysis of the dependence of the expected gamma-ray flux on the DM distribution has been performed in Ref.~\cite{Reis:2024wfy}.}.
Moreover, ${E_{p}^{\rm min}(E_{\gamma})}$ is the minimum CR proton energy required to produce a gamma ray with energy $E_{\gamma}$ \cite{Goncalves:2025nij}, and ${d\Phi_p}/{dE_p}$  is the energy-dependent CR proton flux in the GC region. We assume this flux follows a power-law behavior given by
$d\Phi_p/{dE_p}(E_{\rm p})
= \Phi_0 (E_{\rm p}/10\,\rm TeV)^{\rm -\Gamma}$, with  $\Phi_0 = 4 \times 10^{-8}$ cm$^{-2}$s$^{-1}$ TeV$^{-1}$sr$^{-1}$ and $\Gamma = 2.4$. Following Ref.~\cite{Abramowski:2016mir}, the baseline value of $E_{\rm p}^{\rm max}$ is set to 1 PeV. 

Finally, the differential cross-section, ${d\sigma(p+a \rightarrow p + \gamma)}/{dE_{\gamma}}$, is given by~\cite{Dent:2020qev}:
\begin{widetext}
\begin{eqnarray}
\frac{d\sigma_{p+a \rightarrow p + \gamma}}{dE_{\gamma}} & = & \frac{1}{32 \pi m_a |\vec{p}|^2} \left(\frac{1}{2} \sum_{\rm spins} |{\cal{M}}|^2 \right) \,\, \nonumber \\
 & = & 
\frac{1}{32 \pi m_a |\vec{p}|^2} \Big\{e^2 g_{ap}^2\times
 \Big[-\frac{2(m_p^4 - m_p^2(2m_a^2 + s + u) + su)}{(u-m_p^2)^2}  \nonumber\\
& - & \frac{2(m_p^4 - m_p^2(2m_a^2 + s + u) + su)}{(s-m_p^2)^2} + 
    \frac{4(m_p^4 - 3m_a^2m_p^2 - m_p^2(2s + t))}{(s-m_p^2)(u-m_p^2)} \nonumber \\ & + & 
\frac{4(s - m_a^2)(s+t)}{(s-m_p^2)(u-m_p^2)}\Big]\Big\}\,\,,
\label{eq:alp2}
\end{eqnarray}
\end{widetext}
with $\cal{M}$ being the associated scattering amplitude, $|p|^{2} = E_p^2 - m_p^2$, and the Mandelstam variables are defined as $s = m_p^2 + m_a^2 + 2E_p m_a$, $t = q^2 = m_a^2 - 2E_\gamma m_a$ and $u = 2 m_p^2 + m_a^2 - s - t$. 
Plugging this cross-section in Eq. \ref{eq:flux} allow us to estimate the expected gamma-ray flux for a given set of $(m_a,g_{ap})$ values, or, conversely, to use experimental flux measurement to constrain these parameters.

\section{Methodology}\label{sec3}

\subsection{Observational strategy}
This analysis focuses on computing the expected sensitivity of three ground-based observatories to the gamma-ray signature associated with the IC scattering of CR protons off ALPs in the GC region. We consider one current and two near-future gamma-ray observatories: H.E.S.S., CTAO and SWGO, utilizing publicly available instrument response functions (IRFs) for sensitivity computations. 

H.E.S.S. is an array of five Imaging Atmospheric Cherenkov Telescopes (IACTs) located in the Khomas Highlands of Namibia~\cite{HESS}, providing an ideal location for observing the GC region. The energy-dependent acceptance and residual
background flux for H.E.S.S. observations of the inner halo of the MW are extracted from Ref.~\cite{HESS:2022ygk}. These data represent the most up-to-date H.E.S.S.-like observations of the GC region, obtained through the Inner Galaxy Survey carried out since 2015. We define a region of interest (ROI) of radius 3$^{\circ}$ around the GC, further divided into 27 concentric rings of $0.1^\circ$ width, assuming a homogeneous exposure time of 500 hours. 

The next generation of IACTs, the Cherenkov Telescope Array Observatory (CTAO)~\cite{CTAO}, will consist of sites in both the Northern and Southern Hemispheres. CTAO is designed to detect gamma rays between 20~GeV and 300~TeV, improving flux sensitivity by up to an order of magnitude relative to current facilities, while offering angular resolution improvement by a factor of 2–3 and energy resolution as low as 5\% at TeV energies. Since our target is the GC, we employ the expected performance of the Southern array in its \textit{Alpha} configuration, comprising 14 Medium-Sized and 37 Small-Sized Telescopes. The corresponding 
IRFs are taken from the \textit{publicly available} \texttt{prod5-v0.1} dataset~\cite{CTAprod5}, specifically the \texttt{Performance-prod5-v0.1-South-20deg} configuration. For CTAO we consider a ROI with a radius of $5^\circ$ around the GC and 500 hours of homogeneous observation time, divided into 47 concentric rings.

Finally, we incorporate the Southern Wide-field Gamma-ray Observatory (SWGO)~\cite{SWGO,Conceicao:2023tfb}, a planned wide-field water Cherenkov detector to be built in Chile. SWGO is designed to cover energies from a few hundred GeV to the PeV domain. In this work, we utilize the Strawman design and publicly available IRFs described in Ref.~\cite{SWGOirf}. The exposure in this case consists of 10 years of observation time, covering a 10$^\circ$ circular ROI around the GC, divided into 49 concentric annuli of $0.2^\circ$ width.
\begin{figure}[t]
    \centering
    \includegraphics[width=0.8\linewidth]{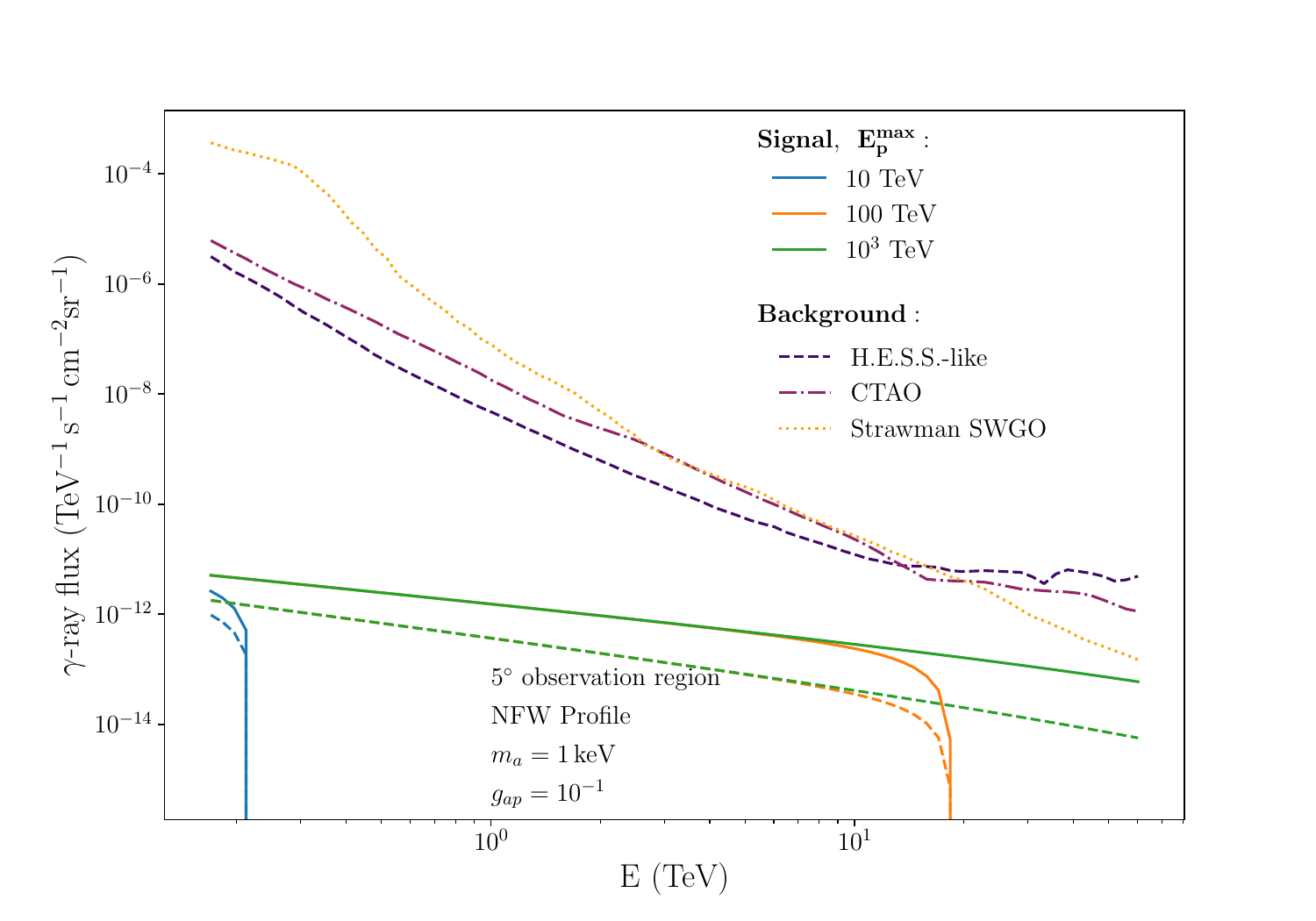}
    \caption{Energy differential gamma-ray flux for the expected ALP-CRp IC scattering signal, as well as the expected residual background from a $5^\circ$ region around the GC region. Here we assume three different observatories: a H.E.S.S.-like array, the CTAO and the SWGO strawman design \cite{SWGOirf}. The expected fluxes assume an ALP of mass $m_a = 1$ keV.}
    \label{fig:expected_flux}
\end{figure}

A crucial ingredient of any VHE gamma-ray 
sensitivity determination is the characterization of backgrounds. For both IACTs and wide-field water Cherenkov detectors, the dominant background arises from the misidentification of hadronic cosmic rays. While most CR-induced showers can be distinguished from gamma-ray events, a non-discriminated fraction 
remains as the residual background. Since no anisotropy has been detected in the VHE CR flux to date, this background is assumed to be spatially isotropic~\cite{Fermi-LAT:2017vjf}. It can be estimated using blank-field observations in extragalactic regions~\cite{HESS:2016mib} or from detailed Monte Carlo simulations of the detector response~\cite{BERNLOHR2013171}. In our analysis, the residual background is derived directly from the IRFs of each observatory. Figure \ref{fig:expected_flux} shows the expected background as seen by CTAO, SWGO and a H.E.S.S.-like observatory as coming from a 5$^\circ$ region around the GC.

In addition to instrumental backgrounds, the GC contains several astrophysical VHE gamma-ray sources and diffuse structures, including the central source HESS J1745–290~\cite{Aharonian:2009zk}, the supernova remnant HESS J1745–303~\cite{Aharonian:2008gw}, and a diffuse component correlated with dense molecular clouds in the Central Molecular Zone~\cite{Abramowski:2016mir}, as well as other sources in the Galactic plane~\cite{HESS:2018pbp}. Modeling these emissions self-consistently is complex and beyond the scope of the present analysis. Instead, we adopt the standard and robust approach of masking bright and extended VHE gamma-ray structures~\cite{HESS:2022ygk}. Following previous GC analyses, we exclude a band of Galactic latitudes between $\pm$0.3$^\circ$ to minimize contamination from these sources.

\subsection{Statistical analysis}

The expected signal counts \(N^S_{ij}\) in each energy bin \(i\) and spatial bin \(j\) are computed by integrating the energy-differential gamma-ray flux from Eq.~(\ref{eq:flux}), convolved with the energy-dependent acceptance \(A_{\rm eff}^\gamma\) and the energy resolution. The latter is modeled as a Gaussian \(G(E-E')\), where $E$ and $E'$ correspond to the reconstructed and true energies, respectively. 
The resulting expression for \(N^S_{ij}\) is
\begin{eqnarray}
   N^S_{ij} = T_{{\rm obs}, j} \int_{E_i - \Delta E_i /2}^{E_i + \Delta E_i /2} dE \int_{-\infty}^{\infty} dE'\, \frac{d\Phi^S_{ij}}{dE'} (\Delta \Omega_j, E') A_{\rm eff}^{\gamma}(E')\, G(E - E') ,
\label{eq:count}
\end{eqnarray}
where \(T_{{\rm obs}, j}\) is the exposure in the spatial bin \(j\). Expected background counts \(N^B_{ij}\) are calculated using the same expression by replacing the signal flux with the residual background flux.

The sensitivity computation for each observatory is determined using a log-likelihood ratio test statistic (TS), a widely adopted method in gamma-ray analyses~\cite{Abdallah:2016ygi, Abdallah:2018qtu, HESS:2022ygk, Reis:2024wfy}. We employ an ON-OFF strategy in which the ROIs serve as ON regions, and the OFF regions, with the same angular size, are used to estimate the background. The likelihood is binned in both energy and spatial dimensions, and is written as
\begin{eqnarray}
\mathcal{L}_{ij} = \textrm{Pois}[N^S_{ij} + N^B_{ij}, N^{\rm ON}_{ij}] \, \textrm{Pois}[\bar{N}^S_{ij} + \bar{N}^B_{ij}, N^{\rm OFF}_{ij}] ,
\label{eq:likelihood}
\end{eqnarray}
where $\textrm{Pois}(k,\lambda)$ stands for the  Poisson distribution and \(N^{\rm ON}_{ij}\) and \(N^{\rm OFF}_{ij}\) denote the observed events in the ON and OFF regions, respectively. \(N^S_{ij}\) and \(\bar{N}^S_{ij}\) represent the expected signal in these regions, while \(N^B_{ij}\) and \(\bar{N}^B_{ij}\) are the corresponding background contributions. Since the residual background is derived from Monte Carlo simulations, we set \(\bar{N}^S_{ij} = 0\). 

The full likelihood is the product of \(\mathcal{L}_{ij}\) over all bins. For a fixed ALP mass \(m_a\) and DM distribution, the signal amplitude \(N_{ij}^S\) is the only free parameter and depends solely on the coupling constant \(g_{ap}\). The TS is defined as
\begin{equation}
    \text{TS}(m_{a}) = - 2 \log \frac{\mathcal{L}(g_{ap}, m_{a})}{\mathcal{L}(\widehat{g}_{ap}, m_{a})} ,
\label{eq:TS}
\end{equation}
where \(\widehat{g}_{ap}\) maximizes the likelihood for a given mass. In the large-statistics limit, the TS follows a \(\chi^2\) distribution with one degree of freedom. One-sided 95\% upper limits on \(g_{ap}\) are obtained by determining the coupling value above the best fit for which TS = 2.71. We derive sensitivity expressed as $95\%$ mean expected upper limits on the $(g_{ap}, m_a)$ parameter space following the Asimov procedure~\cite{Cowan:2011an} widely used for sensitivity computations (see, \textit{e.g.}, Refs.~\cite{Montanari:2023sln, Reis:2024wfy,Goncalves:2025nij}). In this approach, the expected counts in each bin are taken as the expected observation values, so that no statistical fluctuations are included. The profile likelihood ratio is then evaluated as a function of the signal strength parameter, which in our case is determined by the $g_{ap}$ coupling. This procedure provides the median expected exclusion reach in the absence of signal and avoids the need for computationally expensive Monte Carlo pseudo-experiments.

\begin{figure}[t]
    \centering
    \includegraphics[width=0.8\linewidth]{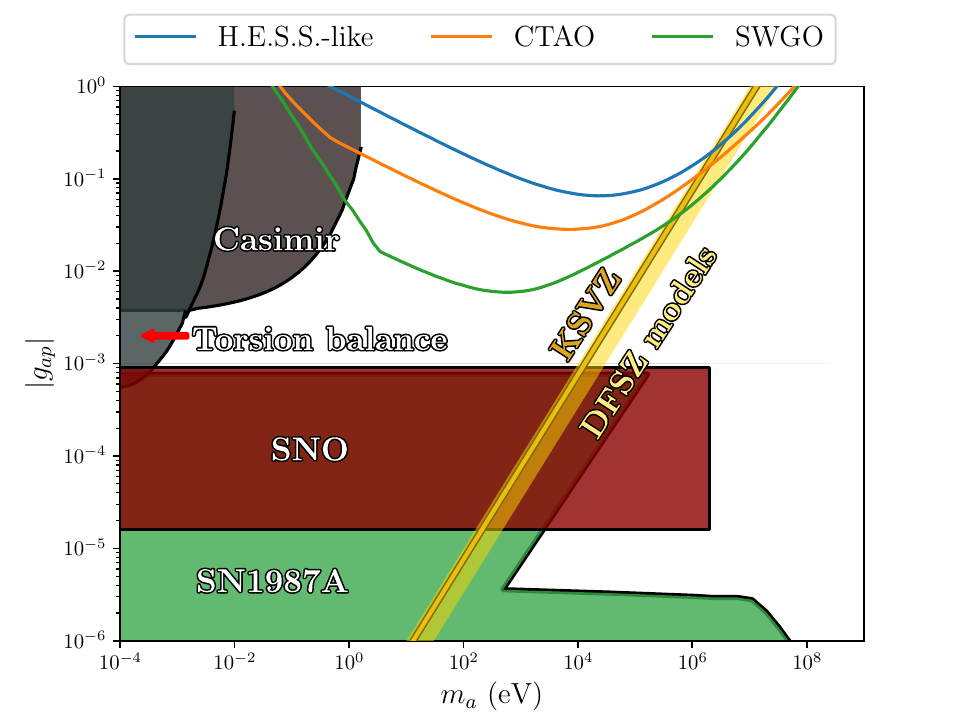}
    \caption{Sensitivity on the ALP-proton coupling $g_{ap}$ versus ALP mass $m_a$. The sensitivity is expressed as 95\% C.L. mean expected upper limits. The sensitivity in the ($m_a$, $g_{ap}$) plane are derived assuming the IC scattering of ALPs and CRp in the GC region, as observed by a H.E.S.S.-like observatory, CTAO and SWGO. We also show exclusion regions from SN1987A observations, SNO, Torsion Balance and Casimir, all extracted from Ref. \cite{AxionLimits}. The golden curve and shaded region represents the portion of the parameter space corresponding to QCD Axions. 
    } 
    \label{fig:results}
\end{figure}

\section{Results and Discussion}\label{sec4}

In Fig.~\ref{fig:results}, we present the 95\% C.L. mean expected upper limits on the ALP-proton coupling $g_{ap}$ as a function of the ALP mass $m_a$, derived from our analysis of H.E.S.S.-like, CTAO, and SWGO potential observations. To provide context for these results, we include current exclusion regions derived from observations of supernova SN 1987A~\cite{Payez_2015}, solar axion searches from the Sudbury Neutrino Observatory (SNO)~\cite{Bhusal_2021}, torsion balance 
experiments~\cite{Kapner_2007,Adelberger_2007}, and measurements of the Casimir 
effect~\cite{Klimchitskaya:2020cnr} (for a compilation of these exclusion regions, see Ref.~\cite{AxionLimits}).

Our results demonstrate that ground-based VHE gamma-ray observatories are capable of probing a vast and currently unexplored region of the $(g_{ap}, m_a)$ parameter space. In fact, in the mass range of $\sim10^{2}$ to $10^{4}$ eV, these observatories can reach a sensitivity down to $g_{ap} \sim 5\times10^{-2}$. In the heavier mass range of $\sim 10^{5} - 10^{8}$ eV the ground-based gamma-ray observatories considered in this study are able to probe the parameter space associated with QCD axion DM in the MeV mass range. The expected sensitivity grows from the H.E.S.S.-like observatory, to CTAO and then To SWGO. This is due to the technological observatory advances, and also due to the fact that SWGO can probe a much larger field-of-view with full-day observation cycle.. The QCD axion provides a natural solution to the strong CP problem, and its coupling to nucleons is a fundamental prediction of specific models. We consider two benchmark scenarios: KSVZ model~\cite{Kim:1979if,Shifman:1979if}, where the axion couples to new heavy quarks, and the DFSZ model~\cite{Zhitnitsky:1980tq,Dine:1981rt}, where the axion couples to standard model quarks at the tree level. 

As shown by the golden shaded region in Fig.~\ref{fig:results}, the predicted couplings for these QCD axion models in the MeV range fall within the sensitivity reach of the ground-based observatories considered here. This is significant because this specific mass-coupling window is notoriously difficult to access via traditional axion haloscopes or stellar cooling arguments. Our analysis shows that the CR-DM scattering channel provides a powerful and independent method to test these foundational BSM predictions.



In summary, we have investigated the sensitivity of current and future VHE gamma-ray ground-based observatories to ALP-proton interactions through the production of secondary gamma rays in the MW halo. We have demonstrated that H.E.S.S., CTAO, and SWGO are expected to reach improved sensitivities on constraints to the $g_{ap}$ coupling relative to existing limits in the $1 - 10^8$~eV mass range. Such results indicate that astrophysical  measurements in the VHE gamma-ray regime are a promising alternative to investigate the ALP - proton couplings, enabling a unique probe of the dark matter sector and the baryonic interactions of QCD axions in the MeV regime.

 \begin{acknowledgments}
 This work was conducted in the context of the CTAO Consortium and the SWGO Collaboration. It has made use of the CTAO instrument response functions provided by the CTAO Consortium and Observatory, and the SWGO instrument response functions provided by the SWGO Collaboration.
 This work is supported by the "ADI 2021" project funded by the IDEX Paris ANR-11-IDEX-0003-02, and by FAPESP, process numbers 2021/02027-0 and 2021/01089-1. A.V. is supported by FAPESP, process number 2024/15560-6, and CNPq grant 309613/2025-6. A.V. and I.R. acknowledge the National Laboratory for Scientific Computing (LNCC/MCTI,  Brazil) for providing HPC resources of the SDumont supercomputer (http://sdumont.lncc.br).
 V.P.G. was partially supported by CNPq, FAPERGS and INCT-FNA (Process No. 408419/2024-5
).
 \end{acknowledgments}

\bibliography{biblio}

\end{document}